\documentclass[runningheads]{llncs}
\usepackage[T1]{fontenc}
\usepackage{subcaption}
\usepackage[justification=centering]{caption}
\usepackage{amsmath}
\usepackage{amssymb}
\usepackage[a4paper]{geometry}
\usepackage{graphicx}
\usepackage{hyperref}
\usepackage{tikz}
\usetikzlibrary{quantikz}
\usepackage{float}
\newcommand{\beq}{\begin{equation}}
\newcommand{\eeq}{\end{equation}}

\begin{document}
\title{Solving the 3D Heat Equation with VQA via Remeshing-Based Warm Starts}
\author{Samuel Donachie\inst{1}\orcidID{0009-0006-7033-2838} \and
Ulysse Remond\inst{1,2}\orcidID{0000-0001-5925-1224} \and
Arthur Mathorel\inst{1}\orcidID{0009-0009-1426-1205}
\and Kyryl Kazymyrenko\inst{1}\orcidID{0000-0001-5429-4348}}%
\authorrunning{S. Donachie et al.}%
\institute{EDF R\&D,  7 Bd Gaspard Monge, 91120 Palaiseau, France \and
LUX, Sorbonne University, 4 Pl. Jussieu, 75005 Paris France}

\maketitle              %
\begin{abstract}
Quantum computing holds great promise for solving classically intractable problems such as linear systems and partial differential equations (PDEs). While fully fault-tolerant quantum computers remain out of reach, current noisy intermediate-scale quantum (NISQ) devices enable the exploration of hybrid quantum-classical algorithms. Among these, Variational Quantum Algorithms (VQAs) have emerged as a leading candidate for near-term applications.
In this work, we investigate the use of VQAs to solve PDEs arising in stationary heat transfer. These problems are discretized via the finite element method (FEM), yielding linear systems of the form Ku=f, where K is the stiffness matrix. We define a cost function that encodes the thermal energy of the system, and optimize it using various ansatz families. To improve trainability and bypass barren plateaus, we introduce a remeshing strategy which gradually increases resolution by reusing optimized parameters from coarser discretizations. Our results demonstrate convergence of scalar quantities with mesh refinement.
This work provides a practical methodology for applying VQAs to PDEs, offering insight into the capabilities and limitations of current quantum hardware.

\end{abstract}
\section{Introduction}
In the current state of quantum computing, fully fault-tolerant quantum computers (FTQCs) remain out of reach \cite{Battistel_2023}. Instead, we operate in the so-called NISQ (Noisy Intermediate-Scale Quantum) era, characterized by quantum devices that are both limited in qubit count and prone to noise. Despite these constraints, NISQ devices can still be leveraged for meaningful tasks \cite{quandela2024quantum}.

As the timeline toward FTQC remains uncertain, both academic and industrial communities are actively exploring use cases that can provide value using today’s imperfect quantum hardware.  Surpassing classical solutions typically requires a large number of qubits ---something that current hardware is not yet capable of. While classical solvers are powerful, they often struggle with scalability and can require extensive computational resources. This challenge, called the curse of dimensionality, refers to the growth in computational complexity as the number of variables and/or degrees of freedom increases. The typical size of challenging industrial simulations reaches $10^6$ degrees of freedom. Quantum computing offers a potential way to overcome this limitation: with n qubits, one can represent $2^n$ states, enabling the compact encoding of high-dimensional vectors and matrices. With 50 qubits, we expect to show one day quantum supremacy on some of those complex problems \cite{supercalculator}. Among the many problems of interest, solving linear equations plays a central role in scientific computing, engineering, and industry. In particular, many physical systems are modeled by Partial Differential Equations (PDEs), which are typically discretized into large linear systems. Efficiently solving these is therefore currently critical for simulations in fields as diverse as fluid dynamics \cite{2014i}, structural mechanics \cite{bathe2006finite}, and electromagnetism \cite{hoole1995finite}. 

One particularly active area is optimization, where quantum computing may already offer practical benefits through the use of Variational Quantum Algorithms (VQAs)\cite{Cerezo_2021}. These hybrid quantum-classical algorithms are well-suited for NISQ devices and have shown promise in addressing linear algebra problems \cite{Sato2021,jaffali2024hdesquantumclassicalhybriddifferential,Liu2020,Cao2012,Colibri}. In this work, we investigate the use of VQAs for solving PDE-derived linear systems for a 3D object, with a focus on designing ansätze that are both expressive and hardware-efficient. 
To this end, we also introduce a cascading remeshing strategy, aimed at improving the convergence and stability of VQA-based linear equation solvers.
The goal of this strategy is to use a coarse quantum representation of a solution, requiring less qubits, as a warm start in a more refined mesh. This procedure creates tailored warm starts for PDE-solving VQAs.

Finally, recognizing that current quantum hardware may not yet support the full execution of such VQAs at scale, we propose to use the ideal final state obtained from our simulated VQA as a benchmark. Since this state corresponds to the solution of a well-defined physical problem (in our case, the 3D heat equation $\Delta u = f$), it provides a concrete, structured target for quantum hardware. We suggest that reproducing such physically meaningful states could serve as a valuable and application-driven benchmark for future quantum devices.

\section{Problem and Method}
\subsection{Problem definition}
Our objective here is to solve the heat equation in 3D : $\Delta u = f$ with $u$ being the temperature field and $f$ the heat source term. One common approach to solve Partial Differential Equations is to use the Finite Element Method (FEM), which discretizes the PDE into a linear system\cite{Sato2021}. For instance, in the case of the heat equation, the system can be written as:
\beq K \vec u =\vec f \eeq
Here, $K$ is the stiffness matrix arising from the FEM discretization \cite{bathe2006finite}. It encodes the relationships between the degrees of freedom in the system and reflects both the geometry of the domain and the material or physical properties involved. In the quantum setting, the solution vector $\vec u$ and the source term (charge vector) $\vec{f}$ are encoded as quantum states $\ket{u}$ and $\ket f$, enabling a compact representation and allowing the use of variational algorithms to approximate the solution. Putting this equation into its variational form leads us to define the energy  $E_c$ of a given trial state $\ket x$:%
\beq E_c(\ket{x})=\frac{1}{2} \bra{x} K \ket{x}- \frac{\braket{x}{f}+\braket{f}{x}}{2}\eeq
We use the energy $E_c$ as a cost function, which is minimized by the solution vector $\ket{u}$.
However, we're working with quantum states, which are normalized, whereas the true solution vector $\vec u$ is not necessarily normalized. We therefore introduce a parameterized quantum state $\ket{\psi(\theta)}$, a norm $r\in\mathbb R$, such as a trial state is expressed as $\ket{x} = r\ket{\psi(\theta)}$. The equation we need to optimize then becomes:
\beq\ket{u}=argmin_{r,\theta} \left( \frac{1}{2}r^2\bra{\psi(\theta)}K\ket{\psi(\theta)} - r\frac{\braket{\psi(\theta)}{f}+\braket{f}{\psi(\theta)}}{2} \right)=\Re(r_{opt}\ket{\psi(\theta_{opt})})\eeq
Analytically minimizing this variational equation with respect to $r$, prior to $\theta$, we obtain : \beq r_{opt}=\frac{\braket{\psi(\theta)}{f} +\braket{f}{\psi(\theta)}}{2\bra{\psi(\theta)}K \ket{\psi(\theta)}}\eeq
Leading us to the final expression of the cost function we use in our VQA \cite{Sato2021}: 
\beq \label{eq:costfun}C(\psi(\theta))=-\frac{1}{8}\frac{(\braket{\psi(\theta)}{f}+\braket{f}{\psi(\theta)})^2}{\bra{\psi(\theta)}K\ket{\psi(\theta)}}\eeq

By finding the state that minimizes this equation, we obtain the solution to our problem. Variational Quantum Algorithms (VQAs) leverage a hybrid quantum-classical loop to perform this minimization: a classical optimizer proposes parameters $\theta$ for a parameterized quantum circuit (ansatz), which is executed on a quantum processor. The resulting state is measured, and the outcome of the cost function $C$ seen in \eqref{eq:costfun} is used by the classical optimizer to update the parameters. Performance depends on each component of the loop: the choice of ansatz, the classical optimizer, and the measurement strategy. {As with classical optimizers and preconditioning, different tradeoffs between the complexities of the quantum computations regarding $K$, $\ket f$ and $\ket u$ are possible}. As a comparatively low number of parameters can be stored to represent large states, QPUs provide a sizeable advantage with regard to the efficient generation and evaluation of high-dimensional states. 
It is also possible to interact with K and $\ket f$ using measurements, scaling polylogarithmically with system size \cite{Liu2020,Ulysse}.
It is worth noting that a high fidelity with the target state $|\braket{\psi(\theta}{u}|^2$ does not necessarily imply convergence in energy \ref{Uniform}. %
\subsection{The choice of the Ansatz}
In a Variational Quantum Algorithm (VQA), an ansatz is a parameterized quantum circuit designed to generate trial quantum states. It serves as a flexible model that is potentially optimized to approximate the solution to a given problem. The expressiveness and structure of the ansatz critically affect the performance and accuracy of the algorithm \cite{Sim_2019}.
Firstly, since one of the main goals of this project is to ensure compatibility with today’s NISQ devices, the ansatz must be designed with hardware constraints in mind. This means using quantum gates with high fidelity and limiting the overall circuit depth.
Typically, ansätze consist of alternating layers of rotation gates and layers of entangling gates. It is well known that entangling gates—especially those involving two or more qubits—are significantly more error-prone than single-qubit gates, with their implementation difficulty increasing rapidly with the number of qubits involved \cite{IQM2qb,IONQ2qb}.

We employ the NLocal class from Qiskit (Qiskit version: 1.4.2) \cite{Qiskit} to construct our ansatz.The circuit is constructed by alternating layers of parameterized rotation and entangling gates, with a final layer of rotation gates. In each entangling layer, a linear chain of CNOTs is applied, where every qubit except the last controls an X gate on the subsequent qubit. This pattern is repeated throughout the circuit, while the final layer contains only rotation gates. When two rotation gates are used, each is applied in a separate layer. Using this Qiskit class, we constructed and compared a limited selection of ansätze, each chosen based on specific criteria relevant to our problem. Since we are addressing a heat diffusion problem, it seems that the target quantum state should be real-valued. All numerical results reported in the next section are obtained from noiseless state-vector simulations where the optimizer provides parameters to our simulated PQC, producing the ideal resulting state. The cost function \eqref{eq:costfun} is then computed from this state, and the optimizer adjusts the parameters accordingly, repeating the process until it found the optimal set. We used the NEWUOA\cite{NEWUOA} gradient-free optimization algorithm  throughout this article.  We studied an inhomogeneous problem to showcase that our model can handle arbitrary boundary conditions. Specifically, we imposed Dirichlet conditions, by fixing the temperature on two faces of the 3D object, and Neumann conditions consisting in a uniform heat flux on an entire third face and an additional flux over half of a fourth face.
Then, using Qiskit's NLocal, we benchmark the following ansätze ---on 6 qubits with 36 parameters--- each %
relevant to a real-valued heat-diffusion solution:

\begin{table}[h]
\vspace{-0.5cm}
\caption{List of Ansätze candidates we considered} 
\centering
\begin{tabular}{|p{1em}|p{5.5em}|p{6em}|p{5em}|p{26em}|}
\hline
&Name&Rotation Layer&Entangling Layer&Comments\\
\hline
1&Ry-only&RY& --- &This baseline illustrates the limitations of an ansatz without entanglement.
\\\hline
2&Ry-Cnot&RY&CNOT&The simplest real-valued ansatz: RY yields purely real amplitudes, and the we chose CNOT as the 2 qubit entangling gate. %
\\\hline
3&Ry-Toffoli&RY&CCNOT&Replaces the CNOT with a more complex three-qubit gate to test whether deeper entanglement outweighs its higher error rate.
\\\hline
4&Ry-1Cnot&RY&CNOT&
Rotation layers: $\mathrm{RY}(\theta)$; exactly one CNOT layer inserted just before the final rotation layer.
Provides sparse entanglement while keeping depth minimal.
\\\hline
5&RxRy-Cnot&RX then RY&CNOT&
Adding RX enlarges the reachable Hilbert space  potentially giving the optimizer better control through a more expressible ansatz \cite{liu2024analysisparameterizedquantumcircuits}, it introduces complex amplitudes.
\\\hline
6&RxRy-Toffoli&RX then RY&CCNOT&
Same expressiveness as above but with a higher-order entangling gate. %
\\\hline
\end{tabular}
\label{tab:ansatz}
\vspace{-0.5cm}
\end{table}
\noindent Let us note that we could have used fewer parameters for Ansatz 1 and 4 (namely 6 and 12, respectively). Given the small number of parameters, this did not affect the results presented below. However, if we had chosen those ansätze, we would have used 6/12 parameters in the later stages.

\section{Results}
\subsection{Cold start}
To compare these ansätze on equal footing, we tuned each circuit so that the total number of trainable parameters was approximately the same, instead of fixing the number of layers. We then ran the optimization 100 times on 6 simulated noiseless qubits, each time starting from random initial parameters $\theta$ generated using a pseudo-random number generator (PCG64, from NumPy version: 1.23.5).
It is worth noting that changing the number of layers or trainable parameters can sometimes have a  drastic impact on the performance of an ansatz. We also note that we let the optimizer run until it could no longer find any improvement, using an absolute tolerance of $10^{-12}$ and a relative tolerance of $10^{-9}$, or until it was forcefully stopped after 2 hours. We can interpret these thresholds as resource limits. This second condition was not triggered until we reached 12 qubits. In order to compare them, we used the accuracy of the energy as a metric of convergence. Energy accuracy is defined as $\frac{C(\ket{\psi(\theta)}) }{C(\ket{u})} \times100\%$  reflecting how closely the optimization converges to the true solution. This metric is designated as "Energy" in the tables.
\begin{table}[h]
\vspace{-0.8cm}
\caption{Ansätze scores on 6 qubits (averaged over 100 2h-optimization runs, each with 36 parameters).} 
\centering
\begin{tabular}{|c|c|c|c|c|c|}
\hline
& Ansatz & Mean Energy (\%) & NEWUOA Iterations & Max Energy (\%) & Min Energy (\%)  \\
\hline
1 & Ry & 88.00 \% ± 0.32 & 452 ± 82 & 91.17 \% & 87.97 \% \\
\hline
2 & Ry-Cnot & 86.7 \% ± 6.7 & 2645 ± 1331 & 96.89 \% & 68.70 \% \\
\hline
3 & Ry-Toffoli & 86.31 \% ± 6.52 & 2430 ± 1203 & 96.70 \% & 71.92 \% \\
\hline
4 & Ry-1Cnot & 86.1 \% ± 11.0 & 1104 ± 525 & 93.38 \% & 50.75 \% \\
\hline
5 & Rx-Ry-Cnot & 93.0 \% ± 4.2 & 3887 ± 1875 & 97.14 \% & 54.33 \% \\
\hline
6 & Rx-Ry-Toffoli & 93.0 \% ± 1.7 & 5793 ± 2720 & 97.17 \% & 88.48 \% \\
\hline
\end{tabular}
\label{tab:1}
\end{table}\\
\noindent Compared to the 6 qubits-discretized target energy all ansatz achieve 85\% and above mean energy accuracy. We observe that multi-qubit entangling gates beyond pairs %
do not provide a noticeable advantage at this scale %
Rather than requiring every run to succeed perfectly, we only need most runs to perform well and at least one to converge very well. Based on this, we can distinguish three performance tiers among the ansätze. On average, using 2 rotation gates seems to help but considering only the maximum accuracy, it doesn't have such an impact. These performance insights can help guide the choice of quantum hardware when deploying the VQA in practice. For the rest of the article, thanks to the scores we obtained, we chose to work with the ansätze  2 and 5. We won't be using Toffoli gates since we can see no clear advantage at this scale, moreover, they perform usually worse than the other existing gates on current hardware. We could also take the optimization time into account by considering the number of iterations required to reach the final result. Adding more gates tends to increase this number, and since the overall waiting time scales with the size of the system, this becomes an important factor to consider. Therefore, we need to be cautious when adding complexity to the circuit.

When examining these results, one might notice that some runs perform quite poorly, with minimum accuracies sometimes dropping to around 50\%. This is due to local minima and/or barren plateaus, where the optimizer gets stuck in regions of vanishing gradients. This accuracy can still be considered high and textbook strategies like a restart could help enhance this score. However, in this case and for our sake, it might not be a major concern, as some runs still manage to perform very well.
The problem arises as the system becomes bigger, requiring more qubits: barren plateaus become more prevalent and harder to avoid. Such random cold start-based runs with 9 to 15 qubits can be observed in Table \ref{tab2}. For 9 qubits we had 108 parameters, for 12 qubits we had 252 parameters and for 15 qubits, we had 330 parameters. We will keep the same number of parameters for the rest of the article.
\begin{table}[h]
\vspace{-0.8cm}
\centering
\caption{Ansätze scores with cold starts (averaged over 100 2h-optimization runs). }
\begin{tabular}{|c|c|c|c|c|c|c|}
\hline
Qubits & Par & Ansatz & Mean Energy (\%) & NEWUOA Iterations & Max Energy (\%) & Min Energy (\%)  \\
\hline
9 & 108 & Rx-Ry-Cnot & 83.3  ± 5.5 & 8506 ± 7313 & 92.9 & 78.5  \\
\hline
9 & 108 &  Ry-Cnot & 23.6  ± 2.4 & 4106 ± 1308 & 29.4  & 17.6 \\
\hline
12& 252&  Rx-Ry-Cnot & 38.0  ± 32 & 11716 ± 4160 & 66.9  & 0.70 \\
\hline
12 & 252 & Ry-Cnot & 4.0 ± 0.7 & 4625 ± 2965 & 5.3  & 2.5  \\
\hline
15 & 330 & Rx-Ry-Cnot & 0.21  ± 0.44  & 5352 ± 714 & 2.71  & 0.02  \\
\hline
15 & 330 & Ry-Cnot & 0.53  ± 0.03 & 4867 ± 796 & 0.57  & 0.25  \\
\hline
\end{tabular}
\vspace{-0.5cm}

\label{tab2}
\end{table}\\
The results show lower convergence when working with a more complex system of 9 qubits especially with the second ansatz Ry-Cnot. Using 12 qubits or more, the optimizer always fails to find a way to optimize the parameters and/or it stops after the 2-hour limit we imposed. We can notice a great difference between the two ansatz for 12 qubits. The Ry-Cnot ansatz showcases very low accuracy as it is twice as deep, and only outputs real states making it less expressible. The largest systems' low accuracy can also be explained by the optimizer getting stuck on a barren plateau and/or in local minima.
In such cases, several classical strategies can accelerate convergence like dual-objective optimization, restarting the optimization process, or a warm start. %
\subsection{Uniform start}
\label{Uniform}
In a warm start, the initial state is already close to the target, which helps the optimizer find better parameters faster. Since we optimize parameters and not quantum states directly, we need to identify both a state near the objective and a parameterization within our ansatz that produces it.
Our first strategy is the \textbf{Uniform start}. The parameters of the first layer set a Hadamard-like transformation, by setting the parameters of $R_y$ gates of the first layer to $\frac\pi2$ and the $R_x$ gates to $0$, while the remaining parameters are initialized randomly from [-0.01,0.01]:
\beq
R_y(\theta) =
\begin{bmatrix}
\cos\left(\frac{\theta}{2}\right) & -\sin\left(\frac{\theta}{2}\right) \\
\sin\left(\frac{\theta}{2}\right) & \cos\left(\frac{\theta}{2}\right)
\end{bmatrix},
\quad
R_x(\theta) =
\begin{bmatrix}
\cos\left(\frac{\theta}{2}\right) & -i\sin\left(\frac{\theta}{2}\right) \\
-i\sin\left(\frac{\theta}{2}\right) & \cos\left(\frac{\theta}{2}\right)
\end{bmatrix},
\quad 
R_y\left(\frac \pi 2\right) \ket 0 = d\ket +
\eeq
This prepares a near to equal superposition $\approx \ket + ^{\otimes n}$, a more structured start than full randomness.
In physical terms, it is like setting all components of the system close to the same temperature, rather than assigning random temperatures to each one. This uniform starting condition can help the optimizer avoid getting lost in barren plateaus by giving it a more meaningful direction to begin with.  %
\begin{table}[h]
\label{tab:tab4}
\vspace{-0.5cm}
\centering
\caption{Ansätze scores with Uniform starts (averaged over 100 2h-optimization runs).}
\begin{tabular}{|c|c|c|c|c|c|c|}
\hline
Qubits& Par & Ansatz & Mean Energy (\%) & NEWUOA Iterations & Max Energy (\%) & Min Energy (\%)  \\
\hline
9 & 108 & Rx-Ry-Cnot & 92.4  ± 0.77 & 5489 ± 3162 & 94.4  & 85.9  \\
\hline
9 & 108 & Ry-Cnot & 92.3  ± 6.2 & 4727 ± 2381 & 94.2  & 33.4  \\
\hline
12 & 252 & Rx-Ry-Cnot & 71.6 ± 23 & 4665 ± 2288 & 91  & 0.32  \\
\hline
12 & 252 &  Ry-Cnot & 83.7  ± 20 & 7409 ± 2583 & 92.2  & 11.1  \\
\hline
15 & 330 & Rx-Ry-Cnot & 46.7  ± 2.90  & 5585 ± 594 & 51  & 35  \\
\hline
15 & 330 & Ry-Cnot & 54.71  ± 5.02 & 4707 ± 240 & 63  & 44  \\
\hline
\end{tabular}
\vspace{-0.5cm}
\label{tab:tab4}
\end{table}

\noindent The results improve with regard to the cold starts. Nevertheless, as the system becomes more complex, even this form of structured initialization begins to lose effectiveness. It still works better than a cold start; for 12 and 15 qubits, most of the runs that were doing well stopped because of the time running out and not because the optimizer couldn't find better parameters. In the 15-qubit case using the Ry–Cnot ansatz, the mean fidelity reaches 95\% even though energy convergence is poor, highlighting that fidelity can be misleading as a performance metric. Moreover, convergences using fidelity disregard the physics of the object, leading to discontinuous states, and therefore cannot be used as a cost function \cite{Ulysse}. It will be useful later on when evaluating the performances of a QPU's output.
Usual warm starts are inherently problem-dependent, since  they leverage specific knowledge about the target state or solution landscape. For instance, in combinatorial optimization, a warm start might be obtained by using a classical heuristic solution mapped into the quantum circuit's parameter space\cite{Egger_2021}.
This start only induces a uniform state, meaning that it merely leverages broad information about the solution's smoothness. That said, it remains valuable in scenarios like ours, where we are primarily concerned with properties such as the temperature of an object—typically a uniform value across the object—rather than a more specialized configuration.

In that sense, Hadamard initialization can be better described as a structured cold start: it provides a more uniform and reproducible starting point than random parameters, but it still lacks the informed guidance that characterizes a true warm start. To overcome this limitation, we introduce the Cascade protocol.
\subsection{The Cascade Protocol}
To choose our ansatz, we compared the performance of various circuits, and all of them performed well on smaller systems (6 qubits). However, as we increase the number of qubits, the Hilbert space grows exponentially, and the optimizer struggles more. The quantum states become more complex, and even the number of parameters becomes harder to manage. In this regime, a warm start is currently one of the only viable strategies to get a VQA to converge effectively. Inspired by recent results on quantum solutions in structural mechanics \cite{Ulysse} obtained by our research group, we propose a novel strategy to improve the scalability of variational quantum algorithms (VQAs) by extending optimized solutions obtained on smaller, tractable systems as structured initializations for larger problems. The central idea is to use the solution of a lower-dimensional instance as a coarse-grained approximation, which can then be embedded and refined within a higher-dimensional Hilbert space.
To illustrate this, consider a 6-qubit system whose 64 computational basis states can be naturally interpreted as a 4×4×4 3D grid: 
\beq \label{eq:u6qb} \ket{u}_{6qb}= \sum\limits_{\bar x,\bar y,\bar z} \ket{z_{coarse}z_{fine}y_{coarse}y_{fine}x_{coarse}x_{fine}},\,\text{where }\bar w=w_{coarse}w_{fine}\in\{0,1\}\times\{0,1\}\eeq
This smaller problem can typically be addressed efficiently either using classical techniques and quantum state preparation, or using a simpler VQA. To extend the solution to a more complex instance ---such as a 9-qubit system representing an 8×8×8 grid--- we construct an initialization that embeds the optimized 6-qubit state into the larger state space. Concretely, each basis state of the smaller system becomes embedded in a local 2×2×2 neighborhood of the larger configuration, effectively increasing the resolution of the space. This hierarchical strategy provides a structured and informed starting point, potentially enhancing convergence and solution quality in the variational optimization. 
This Cascade protocol creates a natural way to scale the problem up while preserving useful structure from the lower-dimensional solution.
To visualize a bit better how it works, here is a remeshing of a 2D grid.

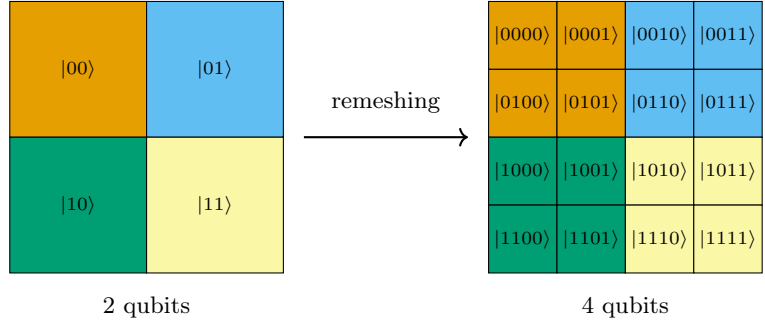
\begin{figure}[h!]
\centering
\begin{tikzpicture}[scale=0.9, every node/.style={minimum size=0.9cm}, on grid]

  \definecolor{quad1}{RGB}{230, 159, 0}   %
  \definecolor{quad2}{RGB}{96, 190, 243}  %
  \definecolor{quad3}{RGB}{0, 158, 115}   %
  \definecolor{quad4}{RGB}{250, 248, 166}  %

  \foreach \x in {0, 1} {
    \foreach \y in {0, 1} {
    
        \ifnum \x = 0
          \ifnum \y = 0
            \def\fillcolor{quad1}
          \else
            \def\fillcolor{quad3}
          \fi
        \else
          \ifnum \y = 0
            \def\fillcolor{quad2}
          \else
            \def\fillcolor{quad4}
          \fi
        \fi
        
      \draw[fill=\fillcolor] (\x*2-2, -\y*2) rectangle ++(2,-2);
      \pgfmathtruncatemacro{\bx}{\x}
      \pgfmathtruncatemacro{\by}{\y}
      \node at (\x*2-1,-\y*2-1) {\scriptsize $\ket{\by\bx}$};
    }
  }
  \node at (0,-4.5) {2 qubits};

    \def\states{
        {0000},{0001},{0010},{0011},
        {0100},{0101},{0110},{0111},
        {1000},{1001},{1010},{1011},
        {1100},{1101},{1110},{1111}
    }
    
    \foreach \state [count=\n from 0] in \states {
        \pgfmathtruncatemacro \x{mod(\n, 4)}
        \pgfmathtruncatemacro \y{int(\n / 4)}
        \ifnum \x < 2
          \ifnum \y < 2
            \def\fillcolor{quad1}
          \else
            \def\fillcolor{quad3}
          \fi
        \else
          \ifnum \y < 2
            \def\fillcolor{quad2}
          \else
            \def\fillcolor{quad4}
          \fi
        \fi
        \draw[fill=\fillcolor] (\x+5,-\y) rectangle ++(1,-1);
        \node at (\x+5.5,-\y-0.5) {\scriptsize $\ket{\state}$};
    }
    \node at (7,-4.5) {4 qubits};

  \draw[->, thick] (2.3,-2) -- (4.7,-2) node[midway, above] {remeshing};

\end{tikzpicture}
\caption{Remeshing from a 2-qubit 2D-grid into a 4-qubit 2D-grid. Each 2-qubit state expands into a structured patch of finer resolution made of $2^2=$ 4 states.}
\vspace{-0.5cm}
\end{figure}
\noindent Since we want to keep the same element geometry (a cube) we need to add 8 new states. In order to obtain those 8 new finite elements per old element, we need to introduce 3 additional qubits, since $2^3=8$. Suppose that an optimized solution has already been obtained for the 6-qubit instance of the problem. To take advantage of this solution when scaling up, we begin by initializing the full 9 qubit system in the state \beq\ket{u}_{6qb} \otimes \ket{000}\eeq
Next, we apply Hadamard gates to each of the 3 newly added qubits. This creates a uniform superposition over all their basis states, yielding:
\beq\ket{u}_{6qb} \ket{+}\ket+\ket+= \ket{u}_{6qb} \otimes \frac{1}{\sqrt{8}} \sum_{i=0}^{7} \ket{i}  \eeq
However, unlike in one-dimensional systems ---where an additional qubit can simply be appended to the end of the register--- in three-dimensional settings, the mapping between basis states and spatial positions is inherently more complex. In particular, the tensor product structure does not directly preserve spatial locality across axes. To ensure that newly added qubits are correctly positioned to encode higher-resolution features along each spatial dimension, it is necessary to introduce a sequence of SWAP gates. These gates effectively rearrange the qubits, allowing the additional degrees of freedom to control fine-grained resolution along the $x$, $y$, and $z$ axes of the enlarged spatial grid. Concretely, we can think of the 6-qubit state as encoding coordinates with pairs of qubits representing coarse and fine positions \eqref{eq:u6qb} as $\bar w=w_{coarse}w_{fine}\in\{0,1\}\times\{0,1\}$.\\ The procedure can be illustrated as $\ket{\bar z}\ket{\bar y}\ket{\bar x}\overset{+3}{\mapsto}\ket{\bar z}\ket{\bar y}\ket{\bar x}\ket{000}\overset{H}{\mapsto}\ket{\bar z}\ket{\bar y}\ket{\bar x}\ket{+}\ket{+}\ket{+}\overset{swap}{\mapsto}\ket{\bar z}\ket{+}\ket{\bar y}\ket{+}\ket{\bar x}\ket{+}.$\\
By carefully swapping, we ensure the new qubits replace the fine control qubits where higher resolution is needed. Nevertheless, this step is unnecessary as you could construct your circuit to leave unused wires in the right spots. Using this technique, we should get a warm start %
by remeshing the solution of a previous optimization with a lower resolution.\\ 
\begin{figure}[h!]
\vspace{-0.5cm}
\centering
\begin{quantikz}[row sep=0.3cm, column sep=0.15cm]
\lstick{$\ket{0}$} & \qw & \qw & \gate[3][0cm]{A(\theta_{\text{opt}})} & \qw & \gate[6][0cm]{\text{SWAP}} & \gate[6][0cm]{B(\theta)} & \qw \\
\lstick{$...$} & \ghost{} & \push{...} & \ghost{} & \push{...} & \qw & \qw & \push{...} \\
\lstick{$\ket{0}$} & \qw & \qw & \qw & \qw & \qw & \qw & \qw \\
\lstick{$\ket{0}$} & \qw & \gate{H} & \qw & \qw & \qw & \qw & \qw \\
\lstick{$\ket{0}$} & \qw & \gate{H} & \qw & \qw & \qw & \qw & \qw \\
\lstick{$\ket{0}$} & \qw & \gate{H} & \qw & \qw & \qw & \qw & \qw \\
\end{quantikz}
\caption{Circuit of the remeshing cascade strategy using previous results $A(\theta_{\text{opt}})$ and superposition expansion via Hadamard gates and swaps.}
\label{fig:remeshing}
\vspace{-0.5cm}
\end{figure}
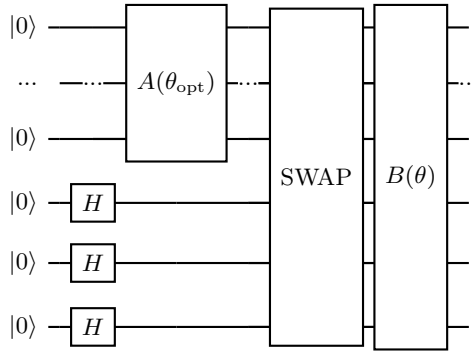

\noindent We define the second ansatz $B(\theta)$ to be the composition of two NLocal ansätze: $Ansatz(\theta) \times Ansatz(0)^\dagger$, such that $B(0) = \mathbb I$. %
If it is possible to implement an identity gate $\mathbb I$ easily with a specific ansatz by simply setting the correct first parameters, this last part isn't needed. However, composing an ansatz in this way simplifies the process, as we only need to initialize all the parameters $\theta$ to $0$ to get $\mathbb{I}$, but obviously complexifies greatly the circuit. We can now compare the results obtained using the Cascade strategy to the Uniform start and the cold start from 9 to 15 qubits on our 3D object. 

Using this strategy, we ran again the optimizations with the two different ansätze. Let's note that the parameters chosen for the first part of the Cascade protocol are the one obtained through a previous optimization on a coarser system. As shown in the previous table, those parameters made us obtain the maximum accuracy recorded for each type of ansatz and were at least 91 \% accurate in energy.\\  
\begin{table}[h]
\label{tab:tab5}
\vspace{-0.5cm}
\centering
\caption{Ansätze scores with the Cascade protocol (averaged over 100 2h-optimization runs).}
\begin{tabular}{|c|c|c|c|c|c|c|}
\hline
Qubits & Par & Ansatz & Mean Energy (\%) & NEWUOA Iterations & Max Energy (\%) & Min Energy (\%)  \\
\hline
6 to 9 & 108 & Rx-Ry-Cnot  & 90.0  ± 0.28 & 4456 ± 1907 & 90.2  & 89.6  \\
\hline
6 to 9 & 108 &  Ry-Cnot & 91.5  ± 0.897 & 10513 ± 955 & 92.0  & 84.6  \\
\hline
9 to 12 &  252 & Rx-Ry-Cnot & 87.4  ± 2.04 & 5209 ± 870 & 88.3  & 66.3  \\
\hline
9 to 12 & 252 &  Ry-Cnot & 87.1  ± 2.37 & 7563 ± 2156 & 90.2  & 80.9  \\
\hline
12 to 15 & 330 & Rx-Ry-Cnot & 72.1 ± 1.45   & 4377 ± 379 & 74.0  & 62  \\
\hline
12 to 15 & 330 & Ry-Cnot & 62.5  ± 4.84 & 3725 ± 892 & 70.3  & 44  \\
\hline
\end{tabular}
\vspace{-0.5cm}
\label{tab:tab5}
\end{table}
The Cascade protocol not only improves convergence but also enhances its consistency, as show by the significantly lower standard deviation of the energy accuracy.
Notably, for the 15-qubit case, the protocol yields higher energy accuracy, successfully achieving our objective of constructing an effective warm-start strategy. Specifically, we observe a 15–20\% (\ref{tab:tab4},\ref{tab:tab5}) improvement compared to the Uniform start, which enables us to meaningfully visualize one of the layers of the target 3D volume \ref{Imaging Cascade}. As expected, at 9 and 12 qubits the overhead of applying an additional layers of gates ---including the cost of possible swapping--- renders the strategy less relevant: the improvement at 9 qubits is negligible, while at 12 qubits the process becomes prohibitively slow relative to the modest gains. At 15 qubits, however, the benefits dominate, with faster convergence and higher accuracy.

\section{Conclusion}
In this work, we explored the use of Variational Quantum Algorithms (VQAs) for solving partial differential equations, focusing on the 3D heat equation discretized via FEM. We applied a noise-free simulation of a VQA to minimize a physically meaningful cost function related to thermal energy. After demonstrating the growing difficulty of such an optimization as system size increases, we introduced the Cascade protocol, a remeshing-based warm-start strategy that reuses optimized parameters from coarser discretizations to initialize larger systems, allowing us to converge with a 15 qubits system. Through extensive simulation, we benchmarked various ansätze and initialization strategies. Basic circuits like Ry-Cnot were %
gradually outperformed by %
more expressive configurations such as Rx-Ry-Cnot
as system size increased. Moreover, since it's using fewer two-qubit gates, the latter are also more noise resistant. The Cascade protocol significantly improved convergence stability and accuracy at larger scales, surpassing both cold and Uniform start baselines. This strategy offers a scalable, physically motivated approach to training VQAs for PDEs, mitigating barren plateaus and improving overall performance.
Moreover, this approach opens the possibility of investigating other initialization schemes, for instance selecting tailored parameter sets at the start of optimization, to evaluate their influence on VQA performance. %

Even if current quantum hardware remains limited, we propose through this article a path towards a 
realistic and structured benchmark for evaluating quantum processing units. By comparing noise-free simulations, noisy models, and real QPU outputs, we highlight both the potential and the current limitations of those devices. In particular, fidelity and energy-based accuracy metrics suggest that small systems can already be meaningfully benchmarked using our approach as seen in \ref{Imaging QPU}.
Moving forward, we envision further improvements through problem-specific warm starts, optimization algorithms, new observables and hardware-aware ansatz design. As quantum hardware matures, the methodology introduced here may serve as a foundation for solving real-world PDEs and also assessing quantum advantage in scientific computing.
\begin{credits}
\subsubsection{\ackname} We would like to thank Joseph Mikael from EDF R\&D for his many fruitful discussions. We are also particularly grateful to Heidi Nelson-Quillin (IonQ) for her valuable technical assistance and the whole IonQ team for their support.
\subsubsection{\discintname}
The authors have no competing interests to declare that are
relevant to the content of this article.
\end{credits}
\bibliographystyle{splncs04}
\bibliography{main}   %
\addcontentsline{toc}{section}{Bibliographie}
\newpage
\appendix
\section{Using our results as a Benchmark}
Building on the results obtained through simulation, we introduce a benchmark strategy for quantum hardware based on physically meaningful quantum states. These states arise from the solution of a 3D heat equation and offer a concrete, interpretable target for evaluating quantum processors.
While executing the full optimization process directly on quantum hardware remains challenging, we propose a more accessible benchmark: given a set of optimized parameters $\theta$, how accurately can a quantum device reproduce the corresponding state $\ket{\psi(\theta)}$? This approach isolates the hardware's ability to generate complex quantum states without requiring full in-loop optimization.
To assess this, we compare the output of the IonQ Aria 1 noise model ---designed to emulate the behavior of the Aria 1 QPU--- to both the ideal solution and the noiseless simulation. Specifically, we evaluate the accuracy with respect to the ideal solution’s energy, the accuracy with respect to the optimized simulation’s energy, and the fidelity between the quantum output and each of these reference states.
As a baseline, we consider the effect of stochastic sampling noise. Using the best parameters previously obtained, we simulate 100 000 measurement shots to produce a reference stochastic state, which allows us to quantify the expected variance in metrics due to finite sampling alone.\\
\begin{table}[h]
\vspace{-0.5cm}
\caption{Simulation results with stochastic noise}
\centering
\begin{tabular}{|c|c|c|c|c|c|c|}
\hline
Qubits & Ansatz & \multicolumn{2}{|c|}{Energy}  & \multicolumn{2}{|c|}{Fidelity}  \\
\hline
&  & stoch/solution & stoch/simulation & stoch/solution & stoch/simulation  \\
\hline
6 & Rx-Ry-Cnot & 93.3  & 99.9  & 96.9  & 99.8  \\
\hline
6 & Ry-Cnot & 97.8  & 99.9  & 98.4  & 99.9  \\
\hline
9 & Rx-Ry-Cnot & 89.3   & 97.2  & 97.8  & 99.8  \\
\hline
9 & Ry-Cnot & 91.2  & 97.0  &  98.8  & 99.9  \\
\hline
12 &  Rx-Ry-Cnot & 40.2   & 46.2  & 97.1  & 98.7  \\
\hline
12 & Ry-Cnot & 42.8   & 47.1  & 97.7  & 98.8  \\
\hline
\end{tabular}
\vspace{-0.3cm}
\end{table}\\
\noindent A clear trend emerges: as the number of qubits increases, the accuracy decreases. Since the simulations are noise-free, this decline can be attributed solely to stochastic sampling noise. The exponential growth of the Hilbert space with the number of qubits leads to a corresponding increase in the number of possible measurement outcomes, requiring more shots to obtain statistically reliable estimates. For example, in the 12 qubit case, the Hilbert space is made of 4096 basis states, and 100,000 shots begin to fall short of delivering accurate statistics across the entire state space. This effect becomes even more pronounced in the 15-qubit case, where the degradation in accuracy due to insufficient sampling is both more severe and more evident. %
\begin{table}[h]
\vspace{-0.5cm}
\centering
\caption{Noisy simulation results}
\begin{tabular}{|c|c|c|c|c|c|}
\hline
Qubits & Ansatz & \multicolumn{2}{|c|}{Energy}  & \multicolumn{2}{|c|}{Fidelity}  \\
\hline
&  & noisy/solution & noisy/simulation & noisy/solution & noisy/simulation  \\
\hline
6 & Rx-Ry-Cnot & 92.4  & 99.0 & 96.8  & 99.6  \\
\hline
6 & Ry-Cnot  & 93.5  & 95.7  & 96.3  & 97.2  \\
\hline
9 & Rx-Ry-Cnot & 80.2   & 87.4  & 96.8  & 98.1  \\
\hline
9 & Ry-Cnot & 64.4  & 68.5  & 92.4  & 93.1  \\
\hline
12 &  Rx-Ry-Cnot & 20.8   & 23.9  & 89.1  & 89.6  \\
\hline
12 & Ry-Cnot & 7.65  & 8.41   &  64.3  & 64.6 \\
\hline
\end{tabular}
\end{table}

\noindent We observe a clear difference between the two ansatz. To match the same number of total parameters, the Rx-Ry-CNOT ansatz uses half as many entangling layers as the Ry-CNOT ansatz. Since two-qubit gates have lower success rates, adding more entangling layers reduces performance, as seen above. Other factors, such as gate types and qubit count, also influence results.
Given hardware costs and simulation limits, we restricted our comparison to 6-qubit circuits with the Ry-CNOT ansatz, running 10,000 shots to obtain the results shown in \ref{Imaging QPU}. \\
\vspace{-0.8cm}
\section{Imaging our results}
\vspace{-0.1cm}
\subsection{The Cascade results}
\label{Imaging Cascade}
As explained above, we used the results obtained through the best optimization thanks to the Cascade warm-start, and applied matplotlib's Lánczos interpolation on the raw statevectors.%
\begin{figure}
    \centering
    \vspace{-0.8cm}
    \begin{subfigure}[b]{0.45\textwidth}
        \centering
        \includegraphics[width=\textwidth]{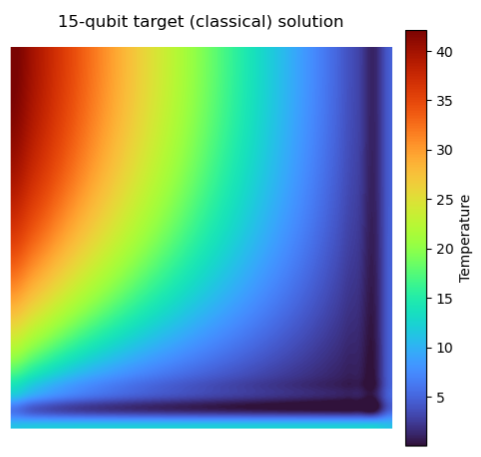}
        
    \end{subfigure}
    \hfill
    \begin{subfigure}[b]{0.45\textwidth}
        \centering
        \includegraphics[width=\textwidth]{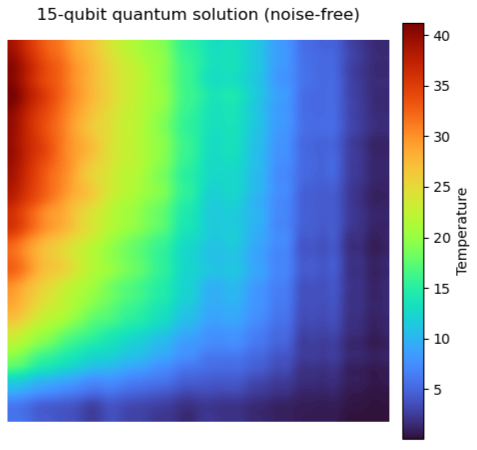}
    \end{subfigure}
    \vspace{-0.2cm}
    \caption{Comparison between the target and the quantum solution for a 15-qubit system.}
    \label{fig:quantum_vs_target}
    \vspace{-0.5cm}
\end{figure}\\
\noindent To visualize this 2D slice of the 3D volume, we used the quantum state obtained from the best optimization run (15 qubits, 330 parameters, Rx-Ry-Cnot Ansatz, with $74\%$ energy accuracy).
We can see how closely the two states match, by noticing the preservation of the general shape and gradients.
\begin{figure}
    \centering
    \includegraphics[height=5.2cm]{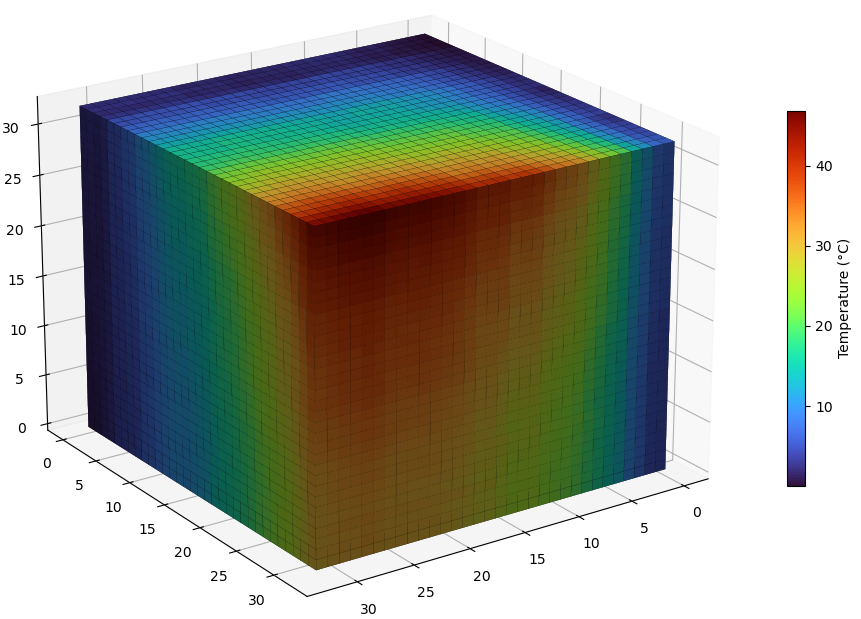}
    \caption{3D Heatmap of the 15 qubits noise-free simulation using the Cascade warm-start}
    \label{fig:enter-label}
\end{figure}

\subsection{Imaging the noisy simulation and QPU's output}
\label{Imaging QPU}

We ran our PQC on IonQ's QPU Aria 1, allowing us to run it against their noise model, the stochastic noise and the noise-free simulation. We retrieved the following results :

\begin{table}
\vspace{-0.5cm}
\centering
\caption{Performance of the Aria 1 noise model and QPU from IonQ for 6 qubits}
\begin{tabular}{|c|c|c|c|c|}
\hline
Metric & Ideal & Stochastic & Noisy (Aria 1) simulation & Aria 1 (QPU) \\
\hline
Energy (state/solution) & 96.8  & 97.1 & 93.5  & 92.2  \\
\hline
Energy (state/simulation) & 100  & 99.3  & 95.7  & 97.8  \\
\hline
Fidelity (state/solution) & 98.6   & 98.2  & 96.3  & 95.3 \\
\hline
Fidelity (state/simulation)  & 100  & 99.8  & 97.2  & 96.3  \\
\hline
\end{tabular}
\end{table}
\noindent These performances show us that all results are close to one another, and both the QPU output and the noisy simulation remain close to the ideal simulated state for such small qubit counts. Naturally, some minor differences and reduced accuracy can still be observed, likely due to noise—whether stochastic or arising from other sources. We can visualize those differences by imaging a slice of our 3D object (made with 6 qubits) as shown previously for 15 qubits.

\begin{figure}[H]
    \centering
    \begin{subfigure}[b]{0.41\textwidth}
        \includegraphics[width=\textwidth]{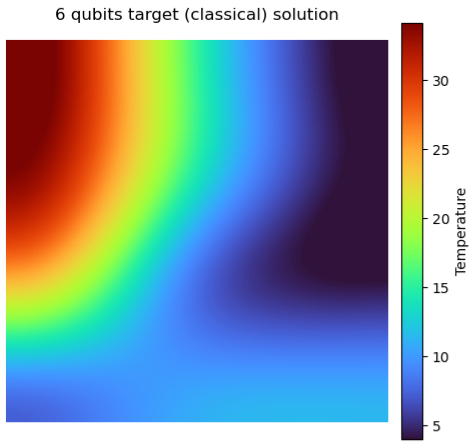}
    \end{subfigure}
    \hfill
    \begin{subfigure}[b]{0.41\textwidth}
        \includegraphics[width=\textwidth]{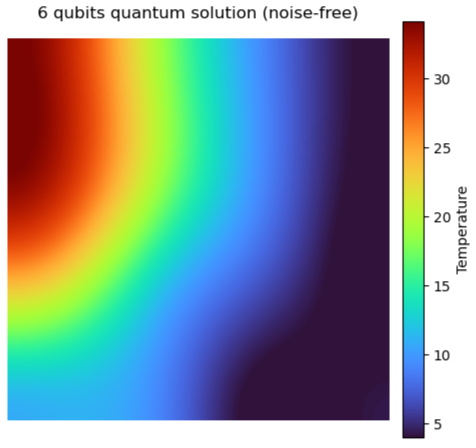}
    \end{subfigure}
    \begin{subfigure}[b]{0.41\textwidth}
        \includegraphics[width=\textwidth]{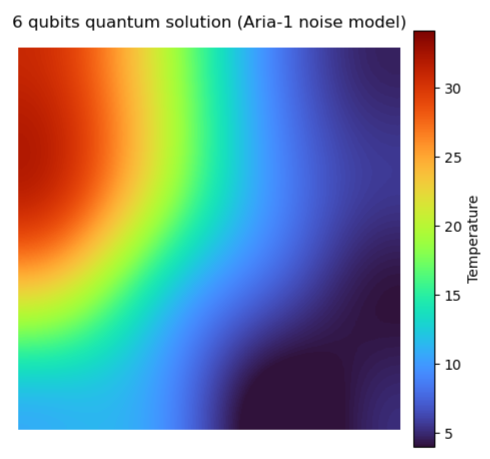}
    \end{subfigure}
    \hfill
    \begin{subfigure}[b]{0.41\textwidth}
        \includegraphics[width=\textwidth]{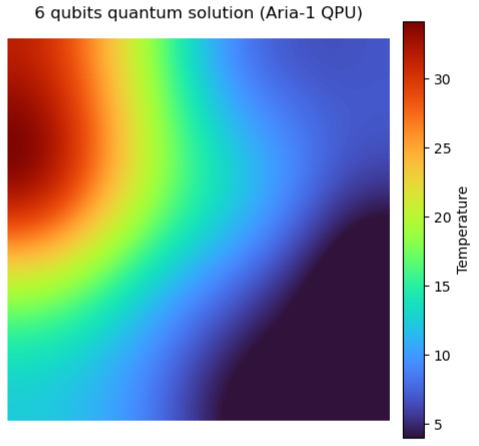}
    \end{subfigure}
    \caption{Comparison between the classical target and quantum solutions for a 6-qubit system using Lánczos interpolation.}
    \label{fig:quantum_vs_target6qb}
\end{figure}
\noindent The noise-free simulation  closely reproduces the overall structure, suggesting that the chosen ansatz captures the essential features of the problem. When introducing the Aria 1 noise model (third), we observe a shift in temperature as observed with the color distribution and the color bar, indicating the degradation caused by noise.
The result from the real QPU (rightmost) exhibits similar features to the noisy simulation but with slightly higher variation, that is likely due to hardware imperfections or shot noise. Despite these limitations, the general shape and gradients are preserved, validating the approach for small systems under realistic condition. 
\end{document}